\documentclass[prb,reprint,amsmath,amssymb,aps]{revtex4-2}

\usepackage{graphicx}
\usepackage{dcolumn}
\usepackage{bm}

\usepackage[utf8]{inputenc}
\usepackage[T1]{fontenc}
\usepackage{mathptmx}

\begin{document}
\title{Plasmon-induced coherence, exciton-induced transparency and Fano interference for hybrid plasmonic systems in strong coupling regime}
\author{Zoe Scott, Shafi Muhammad, and Tigran V. Shahbazyan}
\affiliation{
Department of Physics, Jackson State University, Jackson, MS
39217 USA}


\begin{abstract} 
We present an analytical model describing the transition to a strong coupling regime for an ensemble of emitters resonantly coupled to a localized surface plasmon in a metal--dielectric structure. The response of a hybrid system to an external field is determined by two distinct mechanisms involving collective states of emitters interacting with the plasmon mode. The first mechanism is the near-field coupling between the bright collective state and the plasmon mode which underpins the energy exchange between the system components and gives rise to exciton-induced transparency minimum in scattering spectra in the weak coupling regime and to emergence of polaritonic bands as the system transitions to the strong coupling regime. The second mechanism is the Fano interference between the plasmon dipole moment and the plasmon-induced dipole moment of bright collective state as the hybrid system interacts with the radiation field. The latter mechanism is greatly facilitated by plasmon-induced coherence in a system with the characteristic size below the diffraction limit as the individual emitters comprising the collective state are driven by the same alternating plasmon near field, and therefore, all oscillate in phase. This cooperative effect leads to scaling of the Fano asymmetry parameter and of the Fano function amplitude with the ensemble size, and therefore, it strongly affects the shape of scattering spectra for large ensembles. Specifically, with increasing emitters numbers, the Fano interference leads to a spectral weight shift towards the lower energy polaritonic band.

\end{abstract}
\maketitle

%
\section{Introduction}
\label{sec:intro}

Strong coupling effects \cite{forchel-nature04,khitrova-nphys06,imamoglu-nature06} between surface plasmons in metal-dielectric structures and excitons in J-aggregates \cite{bellessa-prl04,sugawara-prl06,wurtz-nl07,fofang-nl08,guebrou-prl12,bellessa-prb09,schlather-nl13,lienau-acsnano14,shegai-prl15,shegai-nl17,shegai-acsphot19}, in various dye molecules \cite{hakala-prl09,berrier-acsnano11,salomon-prl12,luca-apl14,noginov-oe16}  or in semiconductor nanostructures \cite{vasa-prl08,gomez-nl10,gomez-jpcb13,manjavacas-nl11} have recently attracted intense interest, to a large extent,  due to potential applications including, e.g., quantum computing \cite{waks-nnano16,senellart-nnano17}, ultrafast reversible switching \cite{ebbesen-prl11,bachelot-nl13,zheng-nl16}, and light harvesting \cite{leggett-nl16}. In the strong coupling regime, as the interaction strength between the plasmon and excitons exceeds the losses, the coherent energy exchange between these two components of a hybrid plasmonic system \cite{shahbazyan-nl19} leads to the emergence of polaritonic bands separated by the anticrossing gap (Rabi splitting) \cite{novotny-book}. For a single exciton, the strong coupling transition has been observed only in nanoparticle-on-metal (NoM) systems supporting gap plasmons characterized by extremely strong field confinement \cite{hecht-sci-adv19,pelton-sci-adv19,baumberg-natmat2019}. 
Even prior the transition, in the weak coupling regime, the scattering spectra of hybrid plasmonic systems involving excitons in J-aggregates in colloidal quantum dots (QDs) \cite{haran-nc16,baumberg-nature16,lienau-acsph18,pelton-nc18,pelton-ns19} or  two-dimensional atomic crystals \cite{zhang-nl17,xu-nl17,alu-oe18,urbaszek-nc18,shegai-nl18,zhang-acsphot19,xu-acsphot21} conjugated with Ag or Au nanostructures exhibit a narrow minimum referred to as exciton-induced transparency (ExIT) which has been attributed to the Fano interference between exciton and plasmon excitation pathways \cite{vuckovic-prl06,bryant-prb10,pelton-oe10,agio-prl13} or to the energy exchange imbalance between the system components \cite{shahbazyan-prb20}. For a large number $N$ of excitons coupled to a resonant plasmon mode, the energy exchange takes place between the plasmon and a collective state at a rate proportional to $N$, so that the corresponding coupling parameter $g_{N}$ scales as $g_{N}\propto \sqrt{N}$, leading to very large (up to hundreds meV) observed Rabi splittings  \cite{bellessa-prl04,sugawara-prl06,wurtz-nl07,fofang-nl08,bellessa-prb09,schlather-nl13,lienau-acsnano14,shegai-prl15,shegai-nl17,shegai-acsphot19,hakala-prl09,berrier-acsnano11,salomon-prl12,guebrou-prl12,luca-apl14,noginov-oe16,vasa-prl08,gomez-nl10,gomez-jpcb13,manjavacas-nl11}.

At the same time, the shape of scattering spectra has recently become a subject of considerable interest \cite{savvidis-aom13,ebbesen-fd15,ebbesen-nc15,shegai-nl17,shegai-acsphot19,zhang-nl17,xu-nl17,hughes-optica15,garsia-vidal-njp15,ding-prl17,aizpurua-optica18,xu-acsphot21,shahbazyan-nanophot21}. The widely used classical model of coupled oscillators, which treats the exciton  as "dark" state due to its much smaller (up to $10^{4}$) dipole moment compared to  the plasmon one, predicts a visibly larger spectral weight of the upper polaritonic band that stems from the standard $\omega^{4}$ frequency dependence of the scattering cross-section due to larger re-emission rates by higher-energy states. On the other hand, recent experiments have reported an opposite asymmetry pattern, with the main spectral weight shifted towards the lower polaritonic band, for systems with large exciton numbers \cite{shegai-acsphot19,zhang-nl17,xu-nl17} suggesting that another mechanism, distinct from coherent energy exchange, affects the shape of scattering spectra. Note that the spectrum asymmetry is highly sensitive to the plasmon and QE frequency detuning, and so the change in the asymmetry pattern is clearly observed under precise resonance conditions \cite{shegai-acsphot19}. Recent calculations of absorption and scattering spectra  for a single exciton coupled to plasmon mode \cite{ding-prl17,shahbazyan-nanophot21} pointed at an important role of optical Fano interference, absent in the coupled oscillators model, in shaping the spectra which can lead, for extremely strong field confinement (small plasmon mode volume), to a reversal of the asymmetry pattern. However, such extreme conditions are not normally met in systems involving large exciton numbers $N$, suggesting that some mechanism sensitive to $N$ is at work. Such a mechanism is described in this paper.

Specifically, we develop an analytical model, supported by numerical calculations, for hybrid plasmonic systems comprised of $N$ excitons coupled to resonant plasmon mode which undergoes transition to the strong coupling regime. Starting with the system Hamiltonian with microscopic coupling parameters derived in terms of local fields, we identify the main contributions to the system optical dipole moment induced by the incident electromagnetic  (EM) field. By separating out the bright collective state that is strongly coupled to the plasmon mode, we show that its coupling to the EM field scales with $N$ due to \textit{plasmon-induced coherence}, since the individual dipole moments are all driven by the same alternating plasmon field and, hence, for the system overall size below the diffraction limit, all oscillate \textit{in phase}. As a result, the Fano interference between the dipole moments of the plasmon and  bright state  strongly affects the shape of scattering spectra and, with increasing exciton numbers, can shift the spectral weight towards the lower polaritonic band. This cooperative effect is distinct from  the aforementioned $N$-dependence of the coupling parameter $g_{N}$ as the latter effect originates from coherent energy exchange between the plasmon and the bright collective state while the former results from optical interference between them.

This paper is organized as follows: In Sec.~\ref{sec2}, we outline the system Hamiltonian for coupled excitons and plasmons interacting with the EM field and define the microscopic coupling parameters in terms of local fields. In Sec.~\ref{sec3}, we set up semiclassical Maxwell-Bloch equations for a hybrid plasmonic system and define the dressed plasmon state and the bright and dark collective states, which emerge from the interactions between the plasmon and the ensemble of excitons, as well as the non-diagonal interference contribution to the system induced optical dipole moment. In Sec.~\ref{sec4}, we analyze the ExIT minimum in scattering spectra in terms of energy exchange  between the plasmon and the bright state as well as describe the cooperative Fano interference effect emerging from plasmon-induced coherence. In Sec.~\ref{sec5}, we present the results of numerical calculations of scattering spectra illustrating our findings.

\section{The system Hamiltonian and microscopic coupling parameters}
\label{sec2}

In this section, we outline our general approach to $N$ quantum emitters (QEs) resonantly coupled to a localized plasmon mode in a metal-dielectric structure \cite{shahbazyan-prb21}. For  references, we collect here the standard expressions for the system Hamiltonian  that includes plasmons, QEs, and interactions terms. The plasmon interactions with QEs are defined in terms of normalized local plasmons fields, which determine the plasmon mode volume in a standard way, and the usual relations are established between the Purcell factror and coupling parameters \cite{shahbazyan-nl19}.

We consider $N$ QEs with excitation frequency $\omega_{0}$ and dipole moments $\bm{\mu}_{i}=\mu_{0}\bm{n}_{i}$ situated at positions $\bm{r}_{i}$  ($i=1,\dots,N$) near a metal-dielectric structure characterized by the complex dielectric function  $\varepsilon (\omega,\bm{r})=\varepsilon' (\omega,\bm{r})+i\varepsilon'' (\omega,\bm{r})$ supporting localized plasmon modes with frequencies $\omega_{m}$ interacting with the external  electromagnetic (EM) field $\bm{E}(t)$. We assume that the QE frequency $\omega_{0}$ is close to a plasmon mode frequency $\omega_{m}$ and, in the following,  consider only a single resonant plasmon mode. For the monochromatic EM field of frequency $\omega$, in the rotating wave approximation (RWA), the system dynamics is described by the Hamiltonian 
\begin{align}
\label{H-full}
H=
&\hbar\omega_{m}\hat{a}_{m}^{\dagger}\hat{a}_{m}+\hbar\omega_{0} \sum_{i}\hat{\sigma}_{i}^{\dagger}\hat{\sigma}_{i}+\sum_{i}\hbar g_{im}(\hat{\sigma}_{i}^{\dagger}\hat{a}_{m}+\hat{a}_{m}^{\dagger}\hat{\sigma}_{i})
\nonumber\\
&-\left (\bm{\mu}_{m}\!\cdot\!\bm{E}\,\hat{a}_{m}^{\dagger} e^{-i\omega t} + \sum_{i}\bm{\mu}_{i}\!\cdot\!\bm{E}\,\hat{\sigma}_{i}^{\dagger} e^{-i\omega t} +  {\rm H.c.}\right ),
\end{align}
where $\hat{a}^{\dagger}_{m}$ and $\hat{a}_{m}$ are the plasmon creation and annihilation operators, respectively, $\hat{\sigma}_{i}^{\dagger} $ and $\hat{\sigma}_{i} $ are the raising and lowering operators for QEs, respectively, while the parameters $g_{im}$ and  $\bm{\mu}_{m}$ characterize, respectively, plasmon's coupling to the QE and EM field.

For plasmonic nanostructures with the characteristic size smaller than the radiation wavelength, the coupling parameters can be obtained microscopically by relating them to the system geometry and local field \cite{shahbazyan-prb21}. For such systems, the plasmon modes are determined by the quasistatic Gauss equation \cite{stockman-review}
$\bm{\nabla}\!\cdot\! \left [\varepsilon' (\omega_{m},\bm{r}) \bm{\nabla}\Phi_{m}(\bm{r})\right ]=0$,
where 
$\Phi_{m}(\bm{r})$ is the mode potential, which we choose to be real, that defines the mode field as $\bm{E}_{m}(\bm{r})=-\bm{\nabla}\Phi_{m}(\bm{r})$. In order to determine the plasmon  optical dipole moment, we recast the Gauss's law as $\bm{\nabla}\!\cdot\! \left [\bm{E}_{m}(\bm{r})+4\pi \bm{P}_{m}(\bm{r})\right ]=0$,  where $\bm{P}_{m}(\bm{r})=\chi'(\omega_{m},\bm{r})\bm{E}_{m}(\bm{r})$ is  the electric polarization vector and $\chi=(\varepsilon-1)/4\pi$  is the plasmonic system susceptibility. The plasmon  dipole moment has the form
%
\begin{equation}
\label{mode-dipole}
\bm{p}_{m}=\int dV \bm{P}_{m}=\int dV \chi'(\omega_{m},\bm{r})\bm{E}_{m}(\bm{r}).
\end{equation}
Gauss's equation  does not  determine the overall field normalization \cite{stockman-review}, but the later can be found  by matching the plasmon radiative decay rate and that of a localized dipole with excitation energy $\hbar \omega_{m}$. The plasmon radiative decay rate has the form \cite{shahbazyan-prb18} $\gamma_{m}^{r}=W_{m}^{r}/U_{m}$, where 
\begin{align}
\label{energy-mode}
U_{m}
= \frac{1}{16\pi} 
\!\int \!  dV \, \frac{\partial [\omega_{m}\varepsilon'(\omega_{m},\bm{r})]}{\partial \omega_{m}} \,\bm{E}_{m}^{2}(\bm{r}) ,
\end{align}
is the plasmon mode energy \cite{landau,shahbazyan-prl16} and 
\begin{equation}
\label{power}
W_{m}^{r}=\frac{p_{m}^{2}\omega_{m}^{4}}{3c^{3}},
\end{equation}
is the radiated power  ($c$ is the speed of light)  \cite{novotny-book}. The normalized modes $\tilde{\bm{E}}_{m}(\bm{r})$ are thus determined by setting 
\begin{equation}
\label{rate-mode-rad}
\gamma_{m}^{r}=\frac{4\mu_{m}^{2}\omega_{m}^{3}}{3\hbar c^{3}},
\end{equation}
where $\bm{\mu}_{m}$ is the mode optical transition matrix element. We then find the normalization relation as
\begin{equation}
\label{mode-rescaled}
\tilde{\bm{E}}_{m}(\bm{r})=\frac{1}{2}\sqrt{\frac{\hbar\omega_{m}}{U_{m}}} \bm{E}_{m}(\bm{r}),
\end{equation}
where the scaling factor $\sqrt{\hbar\omega_{m}/U_{m}}$ converts the plasmon  energy $U_{m}$  to $\hbar\omega_{m}$ in order to match the EM field energy (the factor $1/2$ reflects positive-frequency contribution). 
Accordingly, the plasmon optical transition matrix element in the Hamiltonian (\ref{H-full}) takes the form [compare to Eq.~(\ref{mode-dipole})]
\begin{equation}
\label{mode-optical}
\bm{\mu}_{m}=\int dV \chi'(\omega_{m},\bm{r})\tilde{\bm{E}}_{m}(\bm{r}).
\end{equation}
In a similar way, the plasmon non-radiative decay rate is  $\gamma_{m}^{nr}=W_{m}^{nr}/U_{m}$, where $W_{m}^{r}=\frac{1}{8\pi} 
\!\int \!  dV \varepsilon''(\omega_{m},\bm{r})\bm{E}_{m}^{2}(\bm{r})$ is the power dissipated in the plasmonic structure due to Ohmic losses. In terms of normalized fields, the non-radiative rate takes the form
\begin{equation}
\gamma_{m}^{nr}=\frac{1}{2\pi\hbar\omega_{m}} 
\!\int \!  dV \varepsilon''(\omega_{m},\bm{r})\tilde{\bm{E}}_{m}^{2}(\bm{r}),
\end{equation}
and so the plasmon mode's full decay rate is $\gamma_{m}=\gamma_{m}^{nr}+\gamma_{m}^{r}$. Note that in plasmonic structures with a single metallic component, one recovers the standard expression \cite{stockman-review}   $\gamma_{m}^{nr}=2\varepsilon''(\omega_{m})/[\partial\varepsilon'(\omega_{m})/\partial \omega_{m}]$. The optical polarizability tensor of a plasmonic structure describing its response to the external field $\bm{E}e^{-i\omega t}$ has the form 
\begin{equation}
\label{polar-mode}
\bm{\alpha}_{m}(\omega)=\frac{1}{\hbar}\frac{\bm{\mu}_{m}\bm{\mu}_{m}}{\omega_{m}-\omega-\frac{i}{2}\gamma_{m}},
\end{equation}
where we kept only the resonance term \cite{shahbazyan-prb18}.

The QE-plasmon coupling in the Hamiltonian (\ref{H-full}) is expressed via normalized plasmon mode fields as \cite{shahbazyan-prb21}
\begin{equation}
\label{coupling}
\hbar g_{im}=-\bm{\mu}_{i}\!\cdot\!\tilde{\bm{E}}_{m}(\bm{r}_{i}).
\end{equation}
To present the coupling in a cavity-like form, we use the original plasmon mode fields (\ref{mode-rescaled}) to obtain \cite{shahbazyan-nl19}
\begin{equation}
\label{qe-pl-coupling-mode-volume}
g_{im}^{2} = \frac{2\pi \mu_{i}^{2}\omega_{m} }{\hbar {\cal V}_{i}},
~~~
\frac{1}{{\cal V}_{i}}  = \frac{2[\bm{n}_{i}\!\cdot\! \bm{E}_{m}(\bm{r}_{i})]^{2}}{\int \! dV [\partial (\omega_{m}\varepsilon')/\partial \omega_{m}]\bm{E}_{m}^{2}},
\end{equation}
where ${\cal V}_{i}$ is the projected plasmon mode volume characterizing   plasmon field confinement at the QE position $\bm{r}_{i}$ along its dipole orientation $\bm{n}_{i}$ \cite{shahbazyan-prl16,shahbazyan-acsphot17,shahbazyan-prb18}. The plasmon mode volume defines the Purcell factor characterizing radiation enhancement  of a QE near a plasmonic structure:
\begin{equation}
F_{ip}=\frac{\gamma_{i\rightarrow m}}{\gamma_{0}^{r}}=\frac{6\pi c^{3}Q_{m}}{\omega_{m}^{3} {\cal V}_{i}}
\end{equation}
where $Q_{m} =\omega_{m}/\gamma_{m}$ is the plasmon quality factor, $\gamma_{0}^{r}=4\mu_{0}^{2}\omega_{m}^{3}/3\hbar c^{3}$ is the emitter's radiative decay rate (at the plasmon resonance frequency) and $\gamma_{i\rightarrow m}$ is the rate of energy transfer (ET) from the $i$th QE to the plasmon, given by
\begin{equation}
\label{qe-plasmon-rate-mode-volume}
\gamma_{i\rightarrow m}= \frac{8\pi \mu_{0}^{2}Q_{m} }{\hbar {\cal V}_{i}}.
\end{equation}
Comparing Eqs.~(\ref{qe-pl-coupling-mode-volume}) and (\ref{qe-plasmon-rate-mode-volume}), we obtain the relation between the QE-plasmon coupling and decay rates:
\begin{equation}
\label{coupling-rate}
g_{im}^{2}=\frac{1}{4}\gamma_{m}\gamma_{i\rightarrow m}=\frac{F_{ip}}{4}\,\gamma_{m}\gamma_{0}^{r}.
\end{equation}
Thus, all coupling parameters in the Hamiltonian characterizing plasmon interactions with the QE and  EM field are expressed via the system parameters and are related to the plasmon and QE decay rates. Below, we employ these microscopic expressions to elucidate the role of ExIT and Fano interference in the scattering spectra of hybrid plasmonic systems.

\section{Bright and dark collective states and optical dipole moment for a hybrid plasmonic system}
\label{sec3}

We are interested in the linear response of  a system of $N$ QEs resonantly coupled to a   plasmon mode   to the external EM field oscillating at frequency $\omega$ close to resonance with $\omega_{m}$ and $\omega_{0}$. The interactions between QEs and plasmon, described by the QE-plasmon coupling $g_{im}$, give rise to correlations between individual QEs leading, in turn, to the emergence of collective states. If these correlations are dominated by a single resonant plasmon mode, there is just one collective state ("bright" state) that is coupled to the plasmon giving rise, in the strong coupling regime,  to distinct polaritonic bands in the optical spectra. The remaining $(N-1)$ states ("dark" states) are not coupled to the plasmon mode but, nevertheless, can interact with the EM field, which, in contrast to the strongly varying plasmon field, is nearly uniform on the system scale. However, for not very large QE number $N$, the dark states do not contribute significantly to the system optical response due to much smaller  (by several orders) dipole moments of individual QE  relative to the plasmon optical dipole. In this section, we separate out the bright and dark collective state and elucidate their contributions to the system optical dipole moment.

\subsection{Bright and dark collective states}

In the linear response,  there is only a single excitation in the system, either the plasmon or QE and so, for large QE numbers $N$ we can adopt a semiclassical approach by setting up Maxwell-Bloch equations for polarizations $\rho_{i}=\langle\sigma_{i}\rangle$ and $\rho_{m}=\langle a_{m}\rangle$.  Using the Hamiltonian (\ref{H-full}), in the linear approximation, the equations for  $\rho_{m}(t)$  and $\rho_{i}(t)$ are obtained as
\begin{align}
\label{mb}
&i\frac{\partial \rho_{m}}{\partial t}=(\omega_{m}-i\gamma_{m}/2)\rho_{m}+\sum_{i}g_{im}\rho_{i}- (\bm{\mu}_{m}\!\cdot\!\bm{E}/\hbar) e^{-i\omega t},
\nonumber\\
&i\frac{\partial \rho_{i}}{\partial t}=(\omega_{0}-i\gamma_{0}/2)\rho_{i}+g_{im}\rho_{m}- (\bm{\mu}_{i}\!\cdot\!\bm{E}/\hbar)e^{-i\omega t},
\end{align}
where $\gamma_{0}$ is the QE decay rate assumed much smaller than $\gamma_{m}$. Equations (\ref{mb}) define the optical dipole moments of the plasmon and QE  as $\bm{p}_{m}=\bm{\mu}_{m}\rho_{m}$ and $\bm{p}_{i}=\bm{\mu}_{i}\rho_{i}$, respectively, while the hybrid system's response to the EM field is described by its full induced dipole moment $\bm{p}_{s}=\bm{p}_{m}+\sum_{i}\bm{p}_{i}$.

Let us now identify the bright and dark collective states contributing to the system dipole moment $\bm{p}_{s}$. In the steady-state case, substituting $\rho_{m}(t)=\rho_{m}e^{-i\omega t}$ and $\rho_{i}(t)=\rho_{i}e^{-i\omega t}$, the Eq.~(\ref{mb}) takes the form

\begin{align}
\label{mb-ss}
&(\omega-\omega_{m}+i\gamma_{m}/2)\rho_{m}=\sum_{i}g_{im}\rho_{i}+g_{me},
\nonumber\\
&(\omega-\omega_{0}+i\gamma_{0}/2)\rho_{i}=g_{im}\rho_{m}+g_{ie},
\end{align}
where the parameters
\begin{equation}
\label{coupling-em}
 g_{me}=-\bm{\mu}_{m}\!\cdot\!\bm{E}/\hbar,
 ~~~
 g_{ie}=-\bm{\mu}_{i}\!\cdot\!\bm{E}/\hbar,
\end{equation}
characterize the plasmon  and QE   coupling to the external EM field,  respectively. Using the first equation to eliminate $\rho_{m}$ and introducing shorthand notations
\begin{equation}
\label{omega-short}
\Omega_{0}=\omega-\omega_{0}+i\gamma_{0}/2,
~~~
\Omega_{m}=\omega-\omega_{m}+i\gamma_{m}/2,
\end{equation}
the second equation in the system (\ref{mb-ss}) can be written, in operator form, as
\begin{equation}
\label{mb-e}
\Omega_{0}|\rho\rangle =\frac{1}{\Omega_{m}}\left(\hat{G}_{m}|\rho\rangle +g_{me}|g_{m}\rangle \right)+|g_{e}\rangle 
\end{equation}
where we defined the bra and ket vectors in  the QE configurations space as $\langle \rho|=(\rho_{1},\dots,\rho_{N})$, $\langle g_{m}|=(g_{1m},\dots,g_{Nm})$, and $\langle g_{e}|=(g_{1e},\dots,g_{Ne})$, and the coupling operator $\hat{G}_{m}$ with matrix elements $G_{ii'}=g_{im}g_{i'm}$. The bight collective state is defined as the eigenstate of $\hat{G}_{m}$. It is now easy to see that $|g_{m}\rangle$ is, in fact, such an eigenstate  with eigenvalue $g_{N}^{2}$, where 
\begin{equation}
\label{coupling-N}
g_{N}
=\sqrt{g_{1m}^{2}+\dots +g_{Nm}^{2}}
\end{equation}
is the coupling of the bright collective state $|g_{m}\rangle$ with the plasmon  mode, while the rest of eigenstates (dark states) have eigenvalues equal to zero, i.e., they do not couple to the plasmon. 

In order to separate out the contributions of bright and dark collective states, we now introduce the projection operator $\hat{P}=g_{N}^{-2}|g_{m}\rangle\langle g_{m}|$ and decompose the state $|\rho\rangle$ into bright and dark sectors as $|\rho\rangle=|\rho_{b}\rangle+|\rho_{d}\rangle$, where $|\rho_{b}\rangle=\hat{P}|\rho\rangle$ and $|\rho_{d}\rangle=(1-\hat{P})|\rho\rangle$.
Applying operators  $\hat{P}$ and $(1-\hat{P})$ to Eq.~(\ref{mb-e}) and noting that $\hat{G}=g_{N}^{2}\hat{P}$, we obtain
\begin{equation}
\label{rho-bright}
|\rho_{b}\rangle
=\frac{g_{me}+\Omega_{m}g_{N}^{-2}\langle g_{m}|g_{e}\rangle}{\Omega_{m}\Omega_{0}-g_{N}^{2}}|g_{m}\rangle
\end{equation}
and
\begin{equation}
\label{rho-dark}
|\rho_{d}\rangle
= \frac{1}{\Omega_{0}}\left (|g_{e}\rangle-g_{N}^{-2}\langle g_{m}|g_{e}\rangle|g_{m}\rangle\right ).
\end{equation}
From the first line of Eq.~(\ref{mb-ss}), by writing  $\rho_{m}=\Omega_{m}^{-1}\left ( \langle g_{m}|\rho\rangle+g_{me} \right )$ and using the relation $ \langle g_{m}|\rho\rangle= \langle g_{m}|\rho_{b}\rangle$, we obtain 
\begin{equation}
\label{rho-m}
\rho_{m}=\frac{\Omega_{0}g_{me}+\langle g_{m}|g_{e}\rangle}{\Omega_{m}\Omega_{0}-g_{N}^{2}}.
\end{equation}
Finally, introducing the vector $\langle \bm{\mu}|=(\bm{\mu}_{1},\dots,\bm{\mu}_{N})$, the system dipole moment is presented as 
%
\begin{equation}
\label{dipole-full}
\bm{p}_{s}=\bm{\mu}_{m}\rho_{m}+\langle \bm{\mu}|\rho\rangle=\bm{p}_{m}+\bm{p}_{b}+\bm{p}_{d},
\end{equation}
where $\bm{p}_{m}=\bm{\mu}_{m}\rho_{m}$, $\bm{p}_{b}=\langle \bm{\mu}|\rho_{b}\rangle$, and  $\bm{p}_{d}=\langle \bm{\mu}|\rho_{d}\rangle$ are the plasmon, bright and dark  components, respectively. Below, we elucidate the processes contributing to the dipole moment of the hybrid plasmonic system.

\subsection{Optical dipole moment for a hybrid plasmonic system}
Consider first the plasmon component $\bm{p}_{m}=\bm{\mu}_{m}\rho_{m}$ of the system full dipole moment Eq.~(\ref{dipole-full}). According to Eq.~(\ref{rho-m}), it can be decomposed as $\bm{p}_{m}=\bm{p}_{dp}+\bm{p}_{mb}$, where $\bm{p}_{dp}$ is the induced dipole moment of the \textit{dressed plasmon}  and $\bm{p}_{mb}$ is the non-diagonal (interference) term. The former originates from the first term of Eq.~(\ref{rho-m}) and has the form [compare to Eq.~(\ref{polar-mode})]
%
\begin{equation}
\label{dipole-mode}
\bm{p}_{dp}=\frac{1}{\hbar}\frac{\bm{\mu}_{m}(\bm{\mu}_{m}\!\cdot\!\bm{E})}{\omega_{m}+\Sigma_{m}(\omega)-\omega-\frac{i}{2}\gamma_{m}},
\end{equation}
where $\Sigma_{m}(\omega)$ is self-energy term due to plasmon interaction with the bright state which, using Eqs.~(\ref{coupling}) and (\ref{coupling-N}), can be presented as
\begin{equation}
\label{self-mode}
\Sigma_{m}(\omega)
=-\frac{g_{N}^{2}}{\omega_{0}-\omega-\frac{i}{2}\gamma_{0}}
=-\frac{1}{\hbar}\sum_{i}\bm{q}_{i}(\omega)\!\cdot\!\tilde{\bm{E}}_{m}(\bm{r}_{i}).
\end{equation}
Here, we introduced the \textit{plasmon-induced} dipole moment of an individual QE as
\begin{equation}
\label{qe-dipole-induced}
\bm{q}_{i}(\omega)=-\frac{\bm{\mu}_{i}g_{im}}{\omega_{0}-\omega-\frac{i}{2}\gamma_{0}}=\bm{\alpha}_{i}(\omega)\tilde{\bm{E}}_{m}(\bm{r}_{i}),
\end{equation}
where $\bm{\alpha}_{i}(\omega)$ is the QE polarizability tensor, which, near the resonance, is given by
\begin{equation}
\label{polar-qe}
\bm{\alpha}_{i}(\omega)=\frac{1}{\hbar}\frac{\bm{\mu}_{i}\bm{\mu}_{i}}{\omega_{0}-\omega-\frac{i}{2}\gamma_{0}}.
\end{equation}
The imaginary part of plasmon self-energy determines the ET rate from the plasmon to the bright collective state as
\begin{equation}
\label{rate-me}
\gamma_{m\rightarrow b}(\omega)=-2\Sigma''_{m}(\omega)= \frac{ g_{N}^{2} \gamma_{0}}{(\omega-\omega_{0})^{2}+\gamma_{0}^{2}/4}.
\end{equation}
Using Eq.~(\ref{coupling-N}), the above rate can be presented as the sum of individual plasmon-to-QE ET rates, $\gamma_{m\rightarrow b}(\omega)=\sum_{i}\gamma_{m\rightarrow i}(\omega)$, implying collective energy transfer  between the plasmon and the bright collective state \cite{shahbazyan-nl19}. This rate takes its maximal value at resonance $\omega=\omega_{0}$: $\gamma_{m\rightarrow b}=4g_{N}^{2}/\gamma_{0}$.

The non-diagonal contribution to $\bm{p}_{m}$ originates from the second term in the numerator of Eq.~(\ref{rho-m}) and has the form
\begin{equation}
\label{dipole-int-m}
\bm{p}_{mb}=\frac{1}{\hbar}\frac{\bm{\mu}_{m}\left [\bm{q}_{b}(\omega)\!\cdot\!\bm{E}\right ]}{\omega_{m}+\Sigma_{m}(\omega)-\omega-\frac{i}{2}\gamma_{m}},
\end{equation}
where $\bm{q}_{b}(\omega)=\sum_{i}\bm{q}_{i}(\omega)$ is the plasmon-induced dipole moment of the bright  state.
In fact, this  non-diagonal term describes  interference contribution to $\bm{p}_{m}$ by which the dipole moment along $\bm{\mu}_{m}$ is induced by the EM field via interaction with the bright state's plasmon-induced dipole moment $\bm{q}_{b}(\omega)$.

Let us now turn to the bright component $\bm{p}_{b}=\langle \bm{\mu}|\rho_{b}\rangle$, which, according to Eq.~(\ref{rho-bright}), can  be presented as $\bm{p}_{b}=\bm{p}_{bs}+\bm{p}_{bm}$, where $\bm{p}_{bs}$ is induced dipole moment of the bright state itself and $\bm{p}_{bm}$ is the corresponding non-diagonal term. The latter originates from the first term of Eq.~(\ref{rho-bright}) and has the form [compare to Eq.~(\ref{dipole-int-m})] 
\begin{equation}
\label{dipole-int-b}
\bm{p}_{bm}=\frac{1}{\hbar}\frac{\bm{q}_{b}(\omega)\left [\bm{\mu}_{m}\!\cdot\!\bm{E}\right ]}{\omega_{m}+\Sigma_{m}(\omega)-\omega-\frac{i}{2}\gamma_{m}}.
\end{equation}
Note that $\bm{p}_{bm}$ is simply the transpose of   $\bm{p}_{mb}$ with $\bm{q}_{b}(\omega)$ and $\bm{\mu}_{m}$ interchanged. The bright state contribution $\bm{p}_{bs}$ is obtained from the second term of  Eq.~(\ref{rho-bright}) by noting that the bright state's optical matrix element is defined as the projection of $ |\bm{\mu}\rangle$  upon the normalized bright eigenstate, i.e., 
\begin{equation}
\bm{\mu}_{b}=g_{N}^{-1}\langle g_{m}|\bm{\mu}\rangle= \frac{1}{g_{N}} \sum_{i}g_{im}\bm{\mu}_{i},
\end{equation}
so that $\bm{q}_{b}(\omega)=g_{N}\bm{\mu}_{b}/\Omega_{0}$. Then, using the relation $\langle g_{m}|g_{e}\rangle =-g_{N}\bm{\mu}_{b}\!\cdot\!\bm{E}/\hbar$, we obtain
\begin{equation}
\label{dipole-bright}
\bm{p}_{bs}=\frac{1}{\hbar}\frac{\bm{\mu}_{b}(\bm{\mu}_{b}\!\cdot\! \bm{E})}{\omega_{0}+\Sigma_{b}(\omega)-\omega-\frac{i}{2}\gamma_{0}},
\end{equation}
where $\Sigma_{b}(\omega)$ is bright state's self-energy due to its interactions with the plasmon mode [compare to Eq.~(\ref{self-mode})]:
\begin{equation}
\label{self-emitter}
\Sigma_{b}(\omega)=-\frac{g_{N}^{2}}{\omega_{m}-\omega-\frac{i}{2}\gamma_{m}}.
\end{equation}
Similarly, the ET rate from the bright state to the plasmon is
\begin{equation}
\label{rate-em}
\gamma_{b\rightarrow m}(\omega)=-2\Sigma''_{b}(\omega)= \frac{ g_{N}^{2} \gamma_{m}}{(\omega-\omega_{m})^{2}+\gamma_{m}^{2}/4},
\end{equation}
reaching its maximal value at plasmon frequency $\omega_{m}$: $\gamma_{b\rightarrow m}=4g_{N}^{2}/\gamma_{m}$. Using Eq.~(\ref{coupling-N}), $\gamma_{b\rightarrow m}$ can be presented as the sum of individual QE-to-plasmon ET rates, $\gamma_{b\rightarrow m}=\sum_{i}\gamma_{i\rightarrow m}$, where $\gamma_{i\rightarrow m}$  matches Eq.~(\ref{coupling-rate}),  implying collective energy transfer between the bright state and the plasmon \cite{shahbazyan-nl19}. In a narrow frequency interval  $|\omega-\omega_{0}|\lesssim \gamma_{0}$, the plasmon-to-bright-state ET rate $\gamma_{m\rightarrow b}$  \textit{exceeds} the bright-state-to-plasmon ET rate $\gamma_{b\rightarrow m}$,
\begin{equation}
\label{rates-imbalance}
\frac{\gamma_{m\rightarrow b}}{\gamma_{b\rightarrow m}}= \frac{\gamma_{m}}{\gamma_{0}}\gg 1,
\end{equation}
implying \textit{imbalance} between  direct and reverse ET rates in this interval which, in turn, leads to the ExIT minimum in scattering spectra \cite{shahbazyan-prb20}.  Note that the above ratio is independent of the QE number $N$, so the model of ExIT for a single QE \cite{shahbazyan-prb20} can be extended to $N$ QEs, as we show below.

Finally, using Eq.~(\ref{rho-dark}), the dark component $\bm{p}_{d}=\langle \bm{\mu}|\rho_{d}\rangle$ is evaluated in a similar manner as
\begin{equation}
\label{dipole-dark}
\bm{p}_{d}=\frac{1}{\hbar}\sum_{i}\frac{\bm{\mu}_{i}(\bm{\mu}_{i}\!\cdot\! \bm{E})}{\omega_{0}-\omega-\frac{i}{2}\gamma_{0}}
-\frac{1}{\hbar}\frac{\bm{\mu}_{b}(\bm{\mu}_{b}\!\cdot\! \bm{E})}{\omega_{0}-\omega-\frac{i}{2}\gamma_{0}},
\end{equation}
implying that the dark states' contribution to the system dipole moment  is obtained by subtracting  the bright state's spectral weight from that of $N$ isolated QEs.

Thus, the induced dipole moment of a hybrid plasmonic system represents a sum of four distinct contributions,
\begin{equation}
\label{dipole-system2}
\bm{p}_{s}=\bm{p}_{dp}+\bm{p}_{int} +\bm{p}_{bs}+\bm{p}_{d},
\end{equation}
where we combined non-diagonal terms into a single interference contribution,
%
\begin{equation}
\label{dipole-mode-emitter}
\bm{p}_{int}=\bm{p}_{mb}+\bm{p}_{bm}=
\frac{1}{\hbar}\frac{\bm{\mu}_{m}(\bm{q}_{b}\!\cdot\!\bm{E})+\bm{q}_{b}(\bm{\mu}_{m}\!\cdot\!\bm{E})}{\omega_{m}+\Sigma_{m}(\omega)-\omega-\frac{i}{2}\gamma_{m}},
\end{equation}
which is now symmetric in $\bm{\mu}_{m}$ and  $\bm{q}_{b}$. Below, we show that the Fano interference between the plasmon and bright collective state dipole moments, as they both couple to the EM field, can strongly affect the scattering spectra for large $N$ due to plasmon-induced coherence.

\section{Exciton-induced transparency, plasmon-induced coherence and cooperative Fano effect}
\label{sec4}

The four contributions to the optical dipole moment  $\bm{p}_{s}$ [see Eq.~(\ref{dipole-system2})] encode distinct mechanisms by which incident light interacts with a hybrid plasmonic system. The dressed plasmon term $\bm{p}_{dp}$ describes the effect of the plasmonic antenna coupled to QEs via the plasmon near field and provides the most robust features of the optical spectra, such as well-separated polaritonic bands in the strong coupling regime and, prior the transition, the ExIT minimum. The bright state term $\bm{p}_{bs}$ describes the optical response of the collective state coupled to the plasmon mode, which also exhibits  polaritonic bands but, as we demonstrate below, shows no ExIT minimum, in accordance with the energy exchange mechanism of ExIT \cite{shahbazyan-prb20}. Note that the bright state contribution is typically much smaller than the dressed plasmon contribution and therefore has no significant effect on optical spectra. The dark state term  $\bm{p}_{d}$ describes the contribution of collective states that do not couple to the plasmon mode and, thrrefore, only lead to a narrow spectral peak positioned  at the QE frequency. Note that typically we have $\mu_{0}/\mu_{m}\ll 1$, and so for not very large QE number $N$, this contribution is relatively small as well.

In contrast, the non-diagonal term $\bm{p}_{int}$, although smaller that $\bm{p}_{pd}$, can still strongly affect the overall shape of scattering spectra for large QE numbers $N$ via the cooperative Fano interference effect, described below, as the Fano parameter, which determines the shape of scattering spectra, scales with $N$. The underlying mechanism of this effect is \textit{plasmon-induced coherence} in a system with the characteristic size below the diffraction limit as the QEs comprising the bright collective state are all driven by the same alternating plasmon near field and, hence, all oscillate \textit{in-phase} even though the amplitudes may differ  due to plasmon field variations at the QE positions.

\subsection{Exciton-induced transparency: Dressed plasmon vs bright collective  state}

Let us turn to the scattering spectra of a hybrid plasmonic system.  In RWA, the scattering cross-section  $\sigma_{s}^{sc}(\omega)$ is obtained  by normalizing the power $W_{s}=(\omega^{4}/3c^{3})|\bm{p}_{s}(\omega)|^{2}$, radiated  by the system's induced dipole moment  Eq.~(\ref{dipole-system2}),  with the incident flux $S=(c/8\pi)E^{2}$ \cite{novotny-book}. To simplify the analysis, we assume, in this section, that the incident field is polarized along the plasmon dipole moment $\bm{\mu}_{m}=\mu_{m}\bm{n}_{m}$, where $\bm{n}_{m}$ is unit vector, i.e., $\bm{E}=E\bm{n}_{m}$. We also  omit here the contribution of dark states  as they have no significant effect on polaritonic bands as the system transitions to the strong coupling regime, while including it in  numerical calculations in Sec. \ref{sec5}.

We start our analysis with the largest contribution to the system dipole moment coming from the dressed plasmon given by Eq.~(\ref{dipole-mode}). By setting, for a moment, $\bm{p}_{s}\approx\bm{p}_{dp}$, the scattering cross-section takes the form
\begin{equation}
\label{scattering-dp}
\sigma_{dp}^{sc}(\omega)=\frac{8\pi \omega^{4}\mu_{m}^{4}}{3\hbar^{2}c^{4}}
\frac{\left |\omega_{0}-\omega - \frac{i}{2}\gamma_{0}\right |^{2}}
{\left|\left (\omega_{m}-\omega -\frac{i}{2}\gamma_{m}\right )\!\!\left (\omega_{0}-\omega -\frac{i}{2}\gamma_{0}\right )-g_{N}^{2}\right |^{2}}. 
\end{equation}
A similar expression is obtained in the classical coupled oscillator model albeit with phenomenological  parameters \cite{pelton-oe10}, whereas here the coupling  is expressed via the bright-state-to-plasmon ET rate as $g_{N}^{2}=\gamma_{b\rightarrow m}\gamma_{m}/4$. In fact, the microscopic from of coupling parameters is key to interpreting the ExIT  mechanism in terms of energy exchange  \cite{shahbazyan-prb20}, as described by Eq.~(\ref{rates-imbalance}), which we now extend to the case of $N$ QEs.

In the weak coupling regime, the scattering cross-section has the shape of the narrow ExIT minimum sitting on  top of the broad plasmon band. The plasmon scattering cross-section (without QEs) can be obtained either directly from Eq.~(\ref{polar-mode}) or by setting $g_{N}=0$ in Eq.~(\ref{scattering-dp}),
\begin{equation}
\label{scattering-mode}
\sigma_{m}^{sc}(\omega)=
\frac{8\pi \omega^{4}}{3\hbar^{2}c^{4}}\frac{\mu_{m}^{4}}{(\omega_{m}-\omega)^{2}+\gamma_{m}^{2}/4}.
\end{equation}
To trace the emergence of the ExIT minimum, we recast the dressed plasmon scattering cross-section as 
%
\begin{equation}
\label{sigma-sc}
\sigma_{\rm s}^{\rm sc}(\omega)=\sigma_{m}^{\rm sc}(\omega)R_{N}(\omega),
\end{equation}
where the function
\begin{equation}
\label{R-full}
R_{N}(\omega)=\left |\frac{\left (\omega_{m}-\omega-\frac{i}{2}\gamma_{m}\right )\left (\omega_{e}-\omega-\frac{i}{2}\gamma_{e}\right )}
{\left (\omega_{m}-\omega-\frac{i}{2}\gamma_{m}\right )\left (\omega_{e}-\omega-\frac{i}{2}\gamma_{e}\right )-g_{N}^{2}}\right |^{2}
\end{equation}
modulates the plasmon band. In the narrow frequency interval $|\omega_{m}-\omega|/\gamma_{m}\ll 1$, using the relation $g_{N}^{2}=\gamma_{b\rightarrow m}\gamma_{m}/4$, the function  $R_{N}(\omega)$  simplifies to
\begin{equation}
\label{R-weak}
R_{N}(\omega)=\frac{\delta^{2}+1}
{\delta^{2}+(1+p_{N})^{2}},
\end{equation}
where $\delta=2(\omega-\omega_{0})/\gamma_{0}$ is frequency detuning in units of linewidth and the parameter  
\begin{equation}
\label{exitP}
p_{N}=\frac{\gamma_{b\rightarrow m}}{\gamma_{0}}=\frac{4g_{N}^{2}}{\gamma_{m}\gamma_{0}}
\end{equation}
characterizes the  ExIT minimum depth. The ExIT function (\ref{R-weak}) describes the emergence of spectral minimum in the frequency interval $\sim \gamma_{0}(1+p_{N})$ (see numerical results in Sec. \ref{sec5}). 
Note that, with increasing coupling, the ExIT minimum turns into Rabi splitting at $\gamma_{0}p_{N}\sim \gamma_{m}$, recovering the strong coupling transition point $g_{N}\sim \gamma_{m}/2$ (up to factor 1/2).
 
Let us now consider the bright state's contribution to the system dipole moment, given by Eq.~(\ref{dipole-bright}). Omitting all the other terms, i.e., setting $\bm{p}_{s}\approx\bm{p}_{bs}$, the corresponding scattering cross-section is evaluated as 
\begin{equation}
\label{scattering-bs}
\sigma_{bs}^{sc}(\omega)=\frac{8\pi \omega^{4}\mu_{b}^{2}}{3\hbar^{2}c^{4}}
\frac{(\bm{\mu}_{b}\!\cdot\! \bm{n}_{m})^{2}\left |\omega_{m}-\omega - \frac{i}{2}\gamma_{m}\right |^{2}}
{\left|\left (\omega_{m}-\omega -\frac{i}{2}\gamma_{m}\right )\!\!\left (\omega_{0}-\omega -\frac{i}{2}\gamma_{0}\right )-g_{N}^{2}\right |^{2}}.
\end{equation}
In the weak coupling regime, in the narrow frequency interval  $|\omega-\omega_{m}|/\gamma_{m}\ll 1$, we obtain a simple expression
\begin{equation}
\label{scattering-bs2}
\sigma_{bs}^{sc}(\omega)=\frac{8\pi \omega^{4}\mu_{b}^{2}}{3\hbar^{2}c^{4}\gamma_{0}^{2}}
\frac{(\bm{\mu}_{b}\!\cdot\! \bm{n}_{m})^{2}}{\delta^{2}+(1+p_{N})^{2}}
\end{equation}
which has \textit{no} minimum for small values of $\delta$, consistent with the energy exchange mechanism of ExIT discussed above. Note that  relative to the dressed plasmon contribution, the overall magnitude of $\sigma_{bs}^{sc}(\omega)$ is suppressed by a very small factor $\mu_{b}^{4}/\mu_{m}^{4}$, indicating that the bright collective  state  does not, by itself, significantly affect the scattering spectra  (see numerical results in Sec. \ref{sec5}). However, as we show below, the non-diagonal term (\ref{dipole-mode-emitter}) describing the interference effects can strongly affect the overall shape of scattering spectra as the system transitions to the strong coupling regime.

\subsection{Plasmon-induced coherence and cooperative Fano effect}

Let us now include the non-diagonal term (\ref{dipole-mode-emitter}) in the system dipole moment that incorporates coupling of \textit{both} the plasmon and bright state to the EM field.  Since   the incident light is polarized along the plasmon dipole $\bm{\mu}_{m}=\mu_{m}\bm{n}_{m}$, we split $\bm{q}_{b}(\omega)=g_{N}\bm{\mu}_{b}/\Omega_{0}$ in Eq.~(\ref{dipole-mode-emitter}) into parallel and normal components relative to $\bm{n}_{m}$. Then, by setting $\bm{p}_{s}\approx\bm{p}_{dp}+\bm{p}_{int}$, we obtain, after some algebra, the system scattering cross-section, 
\begin{equation}
\label{scattering}
\sigma_{s}^{sc}(\omega)=\frac{8\pi \omega^{4}\mu_{m}^{4}}{3\hbar^{2}c^{4}}
\frac{\left |\omega_{0}+\omega_{F}-\omega - \frac{i}{2}\gamma_{0}\right |^{2}+h^{2}}
{\left|\left (\omega_{m}-\omega -\frac{i}{2}\gamma_{m}\right )\!\!\left (\omega_{0}-\omega -\frac{i}{2}\gamma_{0}\right )-g_{N}^{2}\right |^{2}}, 
\end{equation}
 where $\omega_{F}=-2g_{N}(\bm{\mu}_{m}\!\cdot\!\bm{\mu}_{b})/\mu_{m}^{2}$ is the frequency shift  due to the Fano interference between the plasmon  and bright state's dipole moments and the parameter $h^{2}=g_{N}^{2}[\mu_{b}^{2}-(\bm{n}_{m}\!\cdot\!\bm{\mu}_{b})^{2}]/\mu_{m}^{2}$ characterizes bright state's optical dipole component normal to the plasmon dipole moment. Note, however, that the parameter $h$ is relatively small, since the local fields are strongest along the plasmon oscillation axis, so in the analysis that follows, we set $h=0$ while including it in the numerical results in Sec. \ref{sec5}.

To reveal the role of Fano interference  in  scattering spectra, we recast the  cross-section (\ref{scattering}) as
\begin{equation}
\label{scattering-fano}
\sigma_{s}^{sc}(\omega)=
\sigma_{dp}^{sc}(\omega)F_{N}(\omega),
\end{equation}
%
 where $\sigma_{dp}^{sc}$ is given by Eq.~(\ref{scattering-dp}) and $F_{N}$ is the Fano function: 
\begin{equation}
\label{fano-function}
F_{N}(\omega)= \frac{(\delta-q_{N})^{2}+1}{\delta^{2}+1}.
\end{equation}
Here, $q_{N}=2\omega_{F}/\gamma_{0}$ is the Fano parameter characterizing the scattering spectrum asymmetry which, using Eq.~(\ref{dipole-bright}), can be presented as 
\begin{equation}
\label{fano-QN}
q_{N}
=-\frac{4g_{N}(\bm{\mu}_{m}\!\cdot\!\bm{\mu}_{b})}{\gamma_{0} \mu_{m}^{2}}
=\frac{4   }{\hbar\gamma_{0} \mu_{m}^{2}} \sum_{i}(\bm{\mu}_{m}\!\cdot\!\bm{\mu}_{i})\left [\bm{\mu}_{i}\!\cdot\! \tilde{\bm{E}}_{m}(\bm{r}_{i})\right ].
\end{equation}
The magnitude of $q_{N}$ is determined by the ratio $\mu_{b}/\mu_{m}$, reflecting the fact that Fano interference occurs between the plasmon and bright state dipole moments interacting with the EM field. 

In the case of single QE ($N=1$), the Fano parameter $q_{i}$ takes the form
\begin{equation}
\label{fano-Qi}
q_{i}=-\frac{4g_{im}(\bm{\mu}_{m}\!\cdot\!\bm{\mu}_{i})}{\gamma_{0} \mu_{m}^{2}},
\end{equation}
where, according to Eq.~(\ref{coupling-rate}), the magnitude of individual QE-plasmon coupling $g_{im}$ is related to the Purcell factor $F_{pi}$. The maximal value of $q_{i}$ is achieved when the QE dipole moment is aligned with the plasmon one, $\bm{\mu}_{i}\parallel\bm{\mu}_{m}$, yielding
\begin{equation}
\label{fano-parameters1}
|q_{i}|= \frac{\gamma_{0}^{r}}{\gamma_{0}} \, \sqrt{\frac{F_{pi}}{\eta_{m}}},
\end{equation}
where $\eta_{m}=\gamma_{m}^{r}/\gamma_{m}$ is the plasmon radiation efficiency. Note that the Fano parameter $q_{i}$ is  suppressed by the small factor $\gamma_{0}^{r}/\gamma_{0} \sim 10^{-5}$ due to phonon (or vibron) broadening of  spectral linewidth and, therefore, for single-QE case, it can be appreciable only for small nanostructures  characterized by very strong field confinement with  $F_{pi}\gg 1$ and $\eta_{m}\ll 1$.

For a collective bright state comprised of $N$ QEs, the Fano parameter represents the sum  $q_{N}=\sum_{i}q_{i}$, implying that  $q_{N}$ \textit{scales with the ensemble size}.  This is  a manifestation of plasmon-induced coherence as the plasmon-induced dipole moments of QEs,  comprising the bright state, all oscillate \textit{in phase} since they are driven, in the absence of retardation effects, by the same alternating plasmon near field. As a result, the optical dipole moment of the bright state, given by Eq.~(\ref{dipole-bright}), scales as $\mu_{b}\propto \sqrt{N}$, implying cooperative Dicke-like coupling to the EM field. For the bright state's scattering cross-section (\ref{scattering-bs}), the effect of cooperative coupling is masked 
due to a much larger dressed plasmon contribution, but it does show up in  the  Fano interference, as we demonstrate in the numerical calculations in Sec. \ref{sec5}.

It should be stressed that the cooperative Fano effect persists even for QEs' random dipole orientations.  Indeed, performing orientational averaging in Eq.~(\ref{fano-QN}), we obtain 
\begin{equation}
\label{fano-parameters2}
q_{N}=\frac{4\mu_{0}^{2}N}{3\hbar\gamma_{0}} \,
\frac{\bm{\mu}_{m}\!\cdot\!\tilde{\bm{E}}_{b}}{\mu_{m}^{2}},
\end{equation}
where $\tilde{\bm{E}}_{b}=N^{-1}\sum_{i}\tilde{\bm{E}}_{m}(\bm{r}_{i})$ is the \textit{average} plasmon field acting on  QEs. Note, finally, that the Fano parameter $q_{N}$ is not directly related to the coupling parameter $g_{N}$ as the latter characterizes the energy exchange between the bright state and the plasmon \cite{shahbazyan-nl19} while the former describes the optical interference between them.

%
%
%
%
%
%
%

\section{Numerical results and discussion}
\label{sec5}

\begin{figure}[tb]
\begin{center}
\includegraphics[width=0.95\columnwidth]{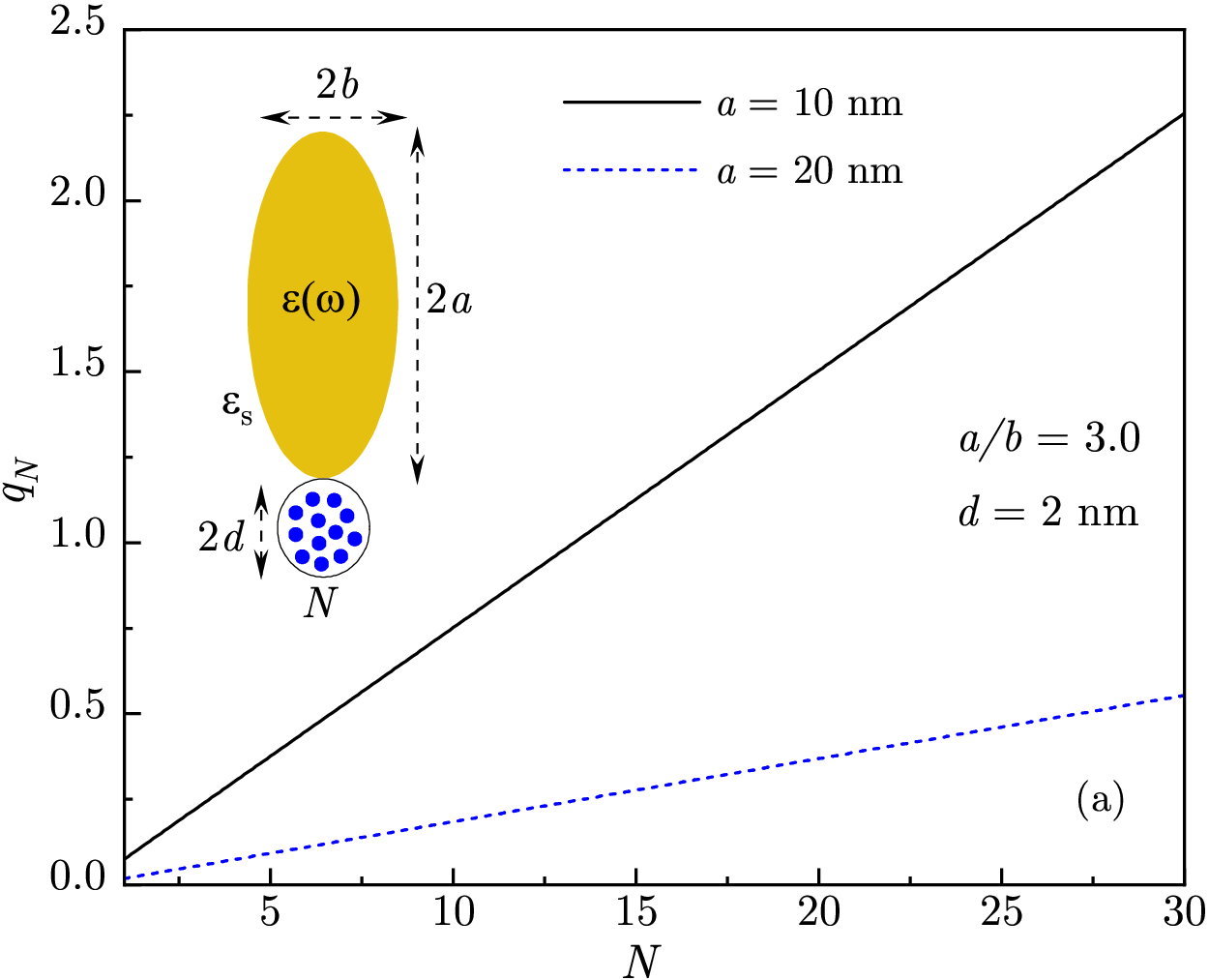}

\vspace{2mm}

\includegraphics[width=0.95\columnwidth]{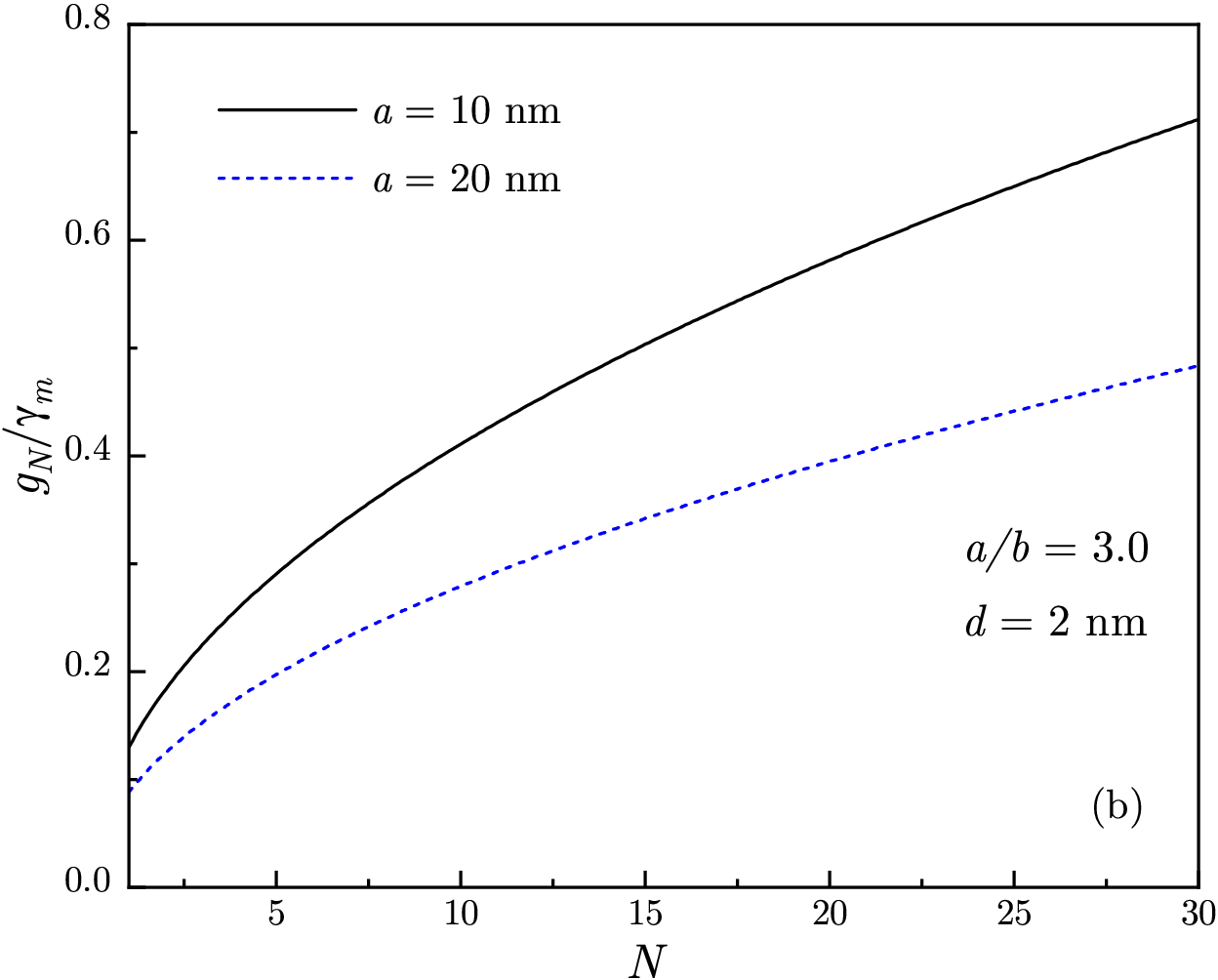}
\caption{\label{fig1} The Fano parameter $q_{N}$ (a) and normalized coupling parameter $g_{N}$ (b) for the bright collective state are plotted against the QE number $N$ for nanorod overall lengths of 40 and 20 nm. Inset: schematics of a hybrid plasmonic system with $N$ QEs distributed within a spherical region of diameter $2d=4$ nm near the tip of an Au nanorod  with the aspect ratio $a/b=3.0$.}
\end{center}
\vspace{-3mm}
\end{figure}

To illustrate our results, below we present the results of numerical calculations for a model system of $N$ QEs placed within a spherical region with radius $d$ near the tip of an Au nanorod in water [see schematics in Fig. \ref{fig1}(a)]. The QE  excitation frequencies were taken in resonance with the surface plasmon frequency, $\omega_{0}=\omega_{m}$, while the QE dipole orientations were chosen to be random. The QE spectral linewidth $\gamma_{0}$ was chosen much smaller than the plasmon decay rate, $\gamma_{0}/\gamma_{m}=0.1$, and the QE radiative decay time was chosen 10 ns, which are typical values for excitons in  quantum dots. For these parameters, the  QE radiative decay rate $\gamma_{0}^{r}$ is much smaller that its spectral linewidth: $\gamma_{0}^{r}/\gamma_{0}\sim 10^{-5}$. The nanorod was modeled by a prolate spheroid with semi-major and semi-minor axes $a$ and $b$, respectively, with the aspect ratio $a/b=3.0$. We used the standard spherical harmonics for a prolate spheroid in order to calculate the local fields to obtain the plasmon parameters  $\mu_{m}$, $\gamma_{m}$, $\eta_{m}$, the coupling parameter $g_{N}$ and the Fano parameter $q_{N}$ which are needed for evaluation of the scattering spectra. In all calculations, have used the Au experimental dielectric function \cite{johnson-christy} and the dielectric constant of water was taken as $\varepsilon_{s}=1.77$. 

%
%

In Fig.~\ref{fig1}, we plot the Fano parameter $q_{N}$ and the coupling between the plasmon and bright collective state $g_{N}$, normalized by the plasmon decay rate $\gamma_{m}$, against the QE number $N$. We carried our calculations for two nanorods with overall lengths 40 and 20 nm in order to control the effect of field confinement (plasmon mode volume) on the shape of scattering spectra. For smaller nanorod ($a=10$ nm), both parameters $q_{N}$ and $g_{N}$ are larger than for the larger nnanorod ($a=20$ nm), although the difference is much greater for the Fano parameter $q_{N}$, which scales linearly with $N$  [see Fig.~\ref{fig1}(a)] vs.  $N^{1/2}$ scaling of $g_{N}$ [see Fig.~\ref{fig1}(b)]. Note that for $N=1$, the system is in the weak coupling regime, while with increasing $N$ both systems transition to the strong coupling regime although the transition points ($g_{N}\approx \gamma_{m}/4$) for them are different.

%
\begin{figure}[tb]
\begin{center}
\includegraphics[width=0.95\columnwidth]{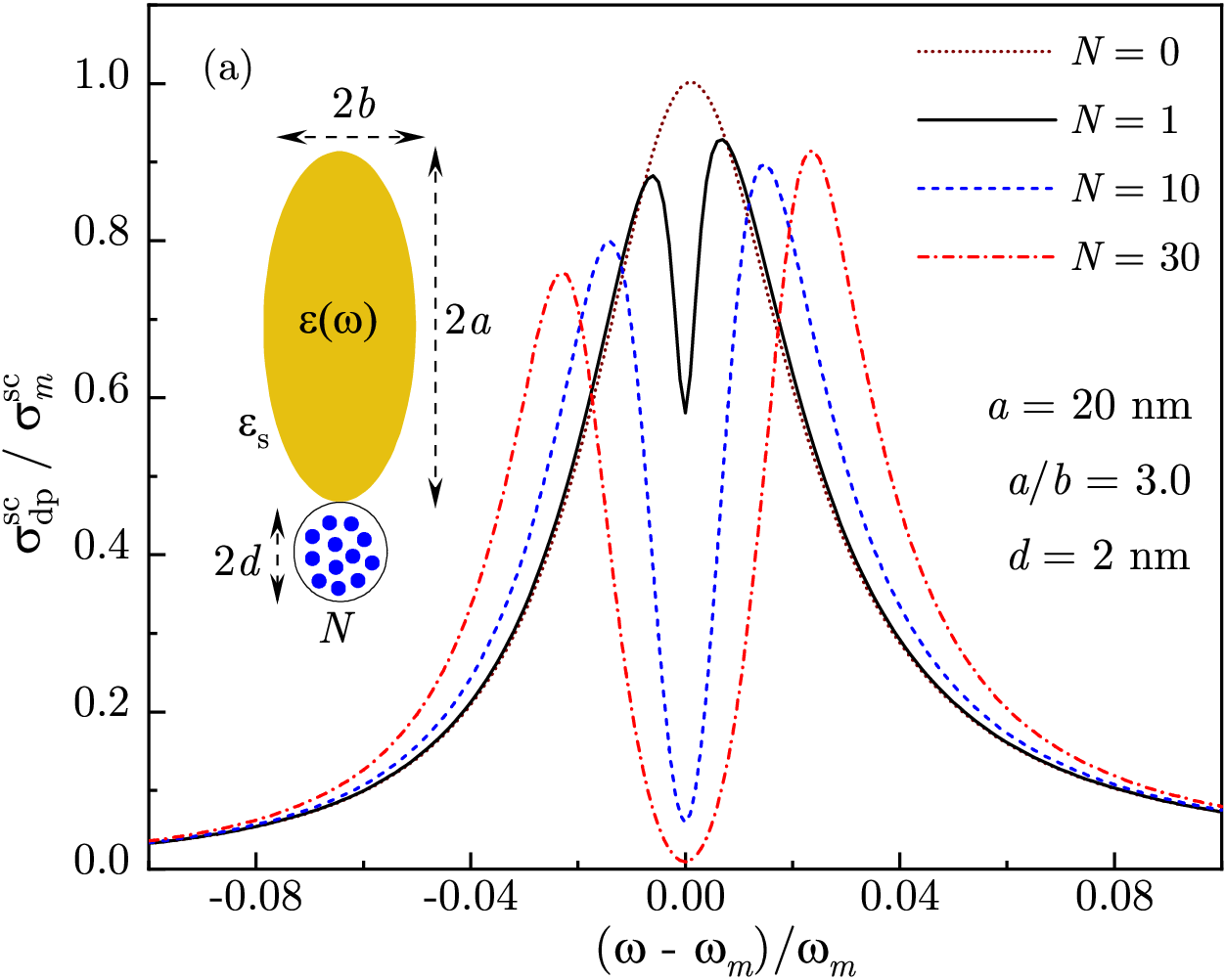}

\vspace{2mm}

\includegraphics[width=0.95\columnwidth]{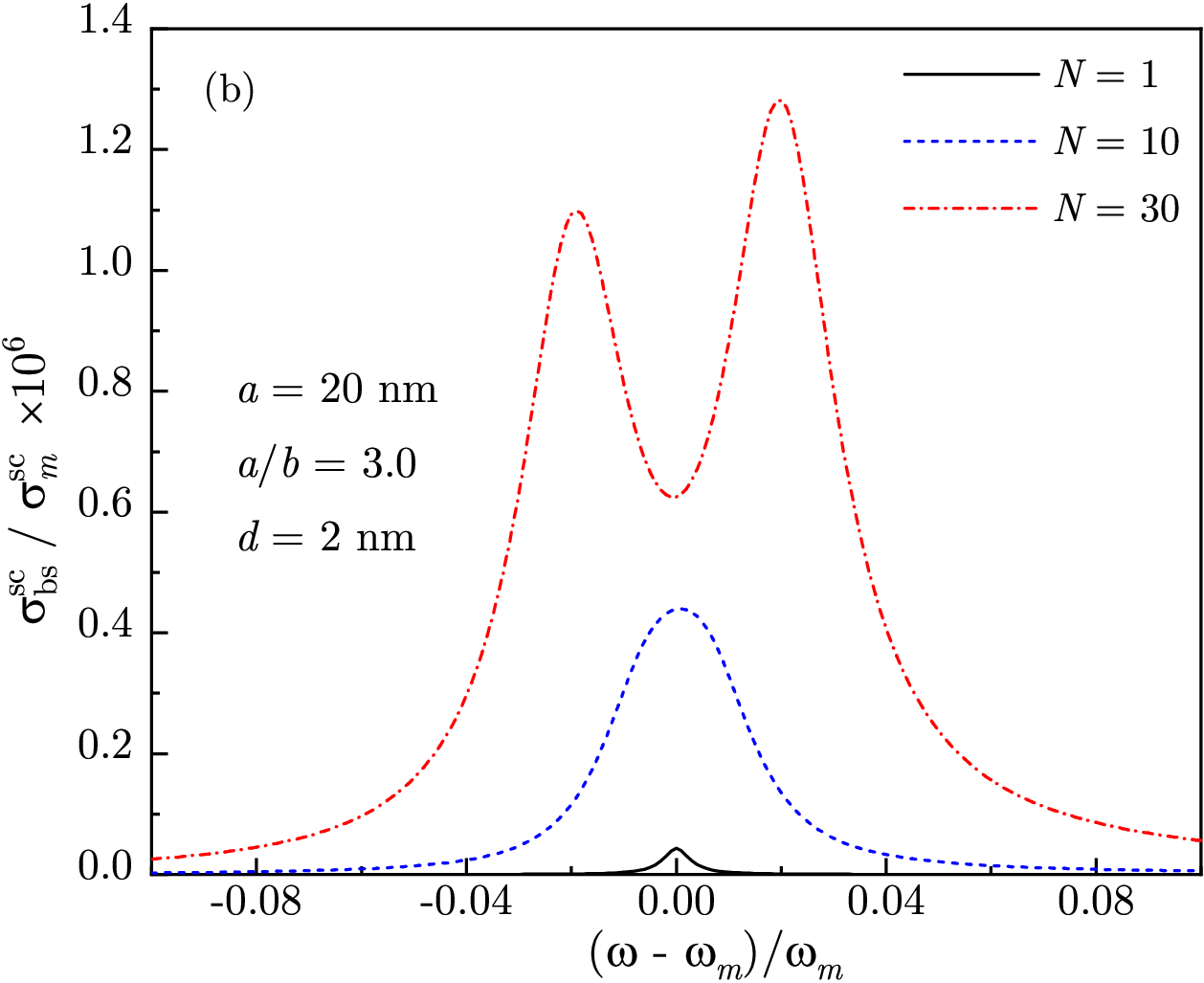}
\caption{\label{fig2} Normalized scattering cross-sections for dressed plasmon (a) and bright collective state (b) are shown for nanorod length 40 nm at several values of QE number $N$. Inset: schematics of a hybrid plasmonic system with $N$ QEs distributed within a spherical region of diameter $2d=4$ nm near the tip of an Au nanorod  with the aspect ratio $a/b=3.0$.
 }
\end{center}
\vspace{-3mm}
\end{figure}
%

%
\begin{figure}[tb]
\begin{center}
\includegraphics[width=0.95\columnwidth]{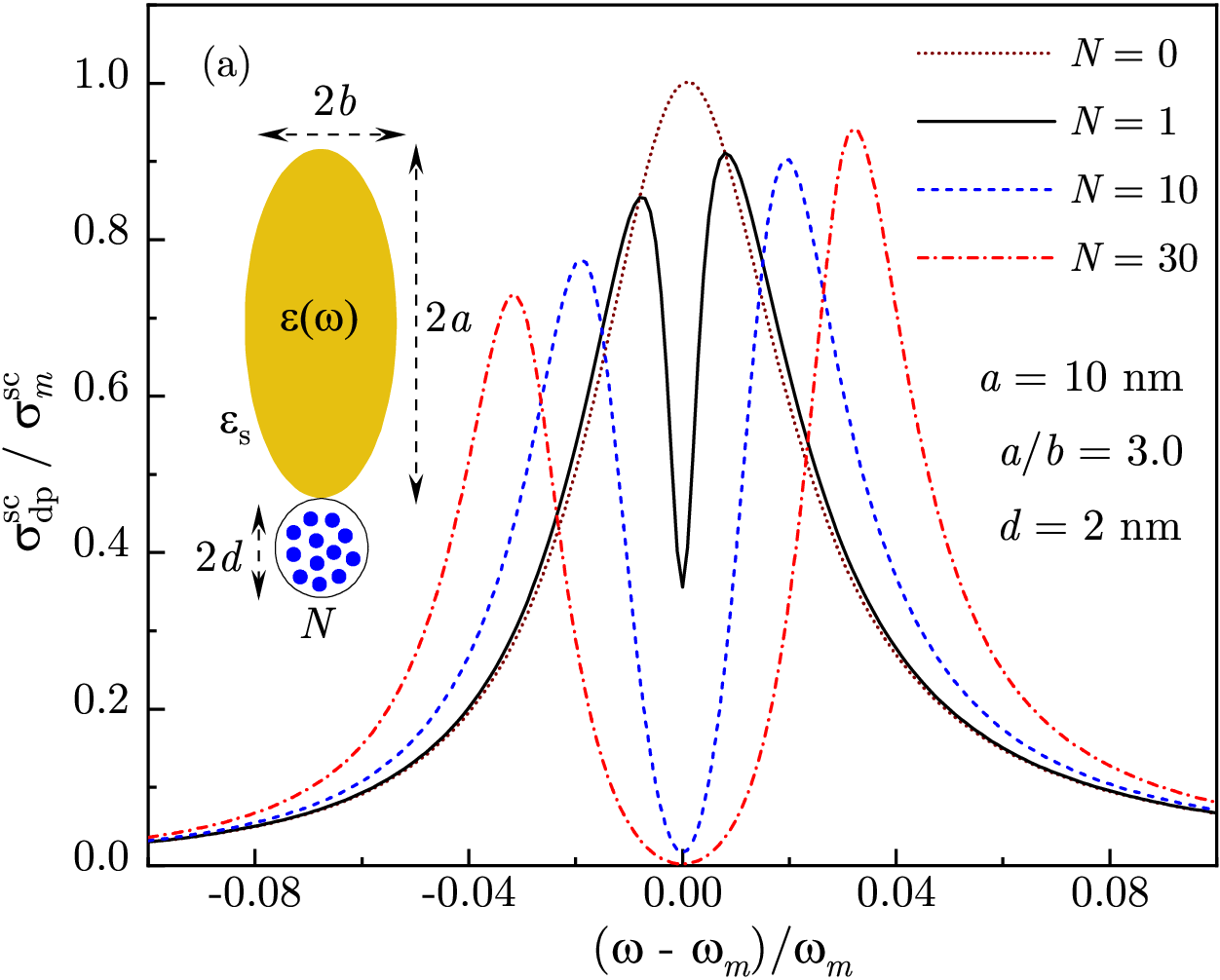}

\vspace{2mm}

\includegraphics[width=0.95\columnwidth]{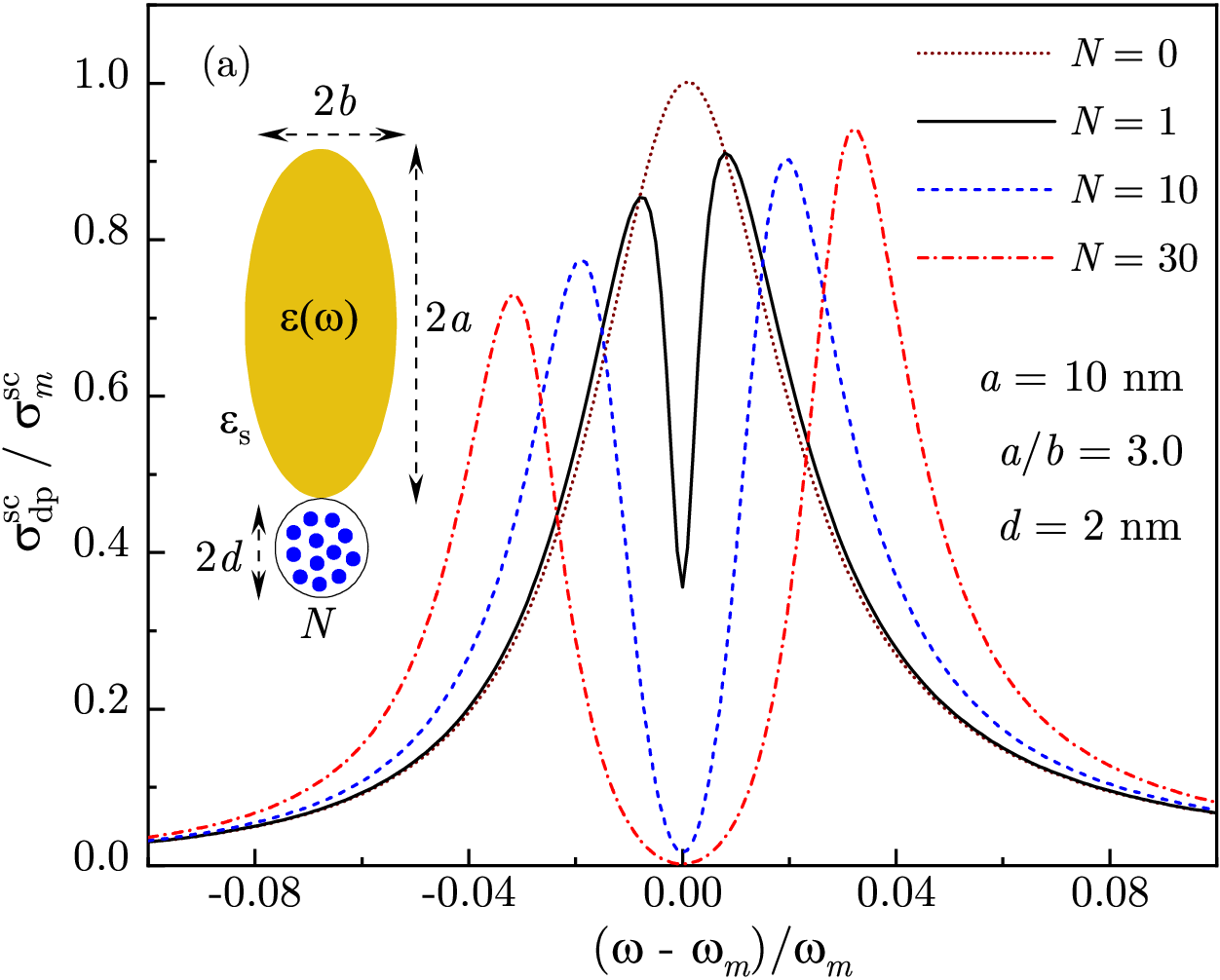}
\caption{\label{fig3} Normalized scattering cross-sections for dressed plasmon (a) and bright collective state (b) are shown for nanorod length 20 nm at several values of QE number $N$. Inset: Schematics of a hybrid plasmonic system with $N$ QEs distributed within a spherical region of diameter $2d=4$ nm near the tip of an Au nanorod  with aspect ratio $a/b=3.0$.
 }
\end{center}
\vspace{-3mm}
\end{figure}
%

In Figs.~\ref{fig2} and \ref{fig3}, we plot the dressed plasmon and bright state scattering scross-sections, but without non-diagonal (interference) contribution, against the incident light frequency for larger and smaller nanorods, respectively. The dressed plasmon spectra, given by Eq.~(\ref{scattering-dp}), are shown in Figs.~\ref{fig2}(a) and \ref{fig3}(a) for several QE numbers $N$ together with the plasmon scattering spectrum ($N=0$), given by Eq.~(\ref{scattering-mode}). In order to maintain a uniform overall scale, we normalize all spectra by the plasmon scattering cross-section at resonance frequency: $\sigma_{m}^{sc}\equiv \sigma_{m}^{sc}(\omega_{m})$. For a single QE ($N=1$), both systems are in the weak coupling regime as their spectra exhibit a narrow ExIT minimum, described by Eq.~(\ref{R-weak}), superimposed upon the broad plasmon band. With increasing $N$, the ExIT minimum transforms into the Rabi splitting as the system transitions to the strong coupling regime. Evidently, the Rabi splitting is more pronounced for the nanorod with  $a=10$ nm [see Fig.~\ref{fig3}(a)] since the transition takes place at smaller $N$ due to stronger field confinement for smaller nanorods [see Fig.~\ref{fig1}(b)].

In Figs.~\ref{fig2}(b) and \ref{fig3}(b) we show the bright state scattering spectra calculated using Eq.~(\ref{scattering-bs}) for otherwise same parameter values. Compared to the dressed plasmon spectra, the bright state spectra are suppressed by the large factor $10^{5}$-$10^{6}$ due to a much greater plasmon dipole moment even for $N=30$. Note that somewhat larger relative amplitudes in Fig.~\ref{fig3}(b) are due to the smaller plasmonic antenna size at $a=10$ nm. Importantly, in the weak coupling regime, the bright state spectra exhibit no ExIT minimum, which is consistent with the energy exchange mechanism of ExIT discussed above.

%
\begin{figure}[tb]
\begin{center}
\includegraphics[width=0.95\columnwidth]{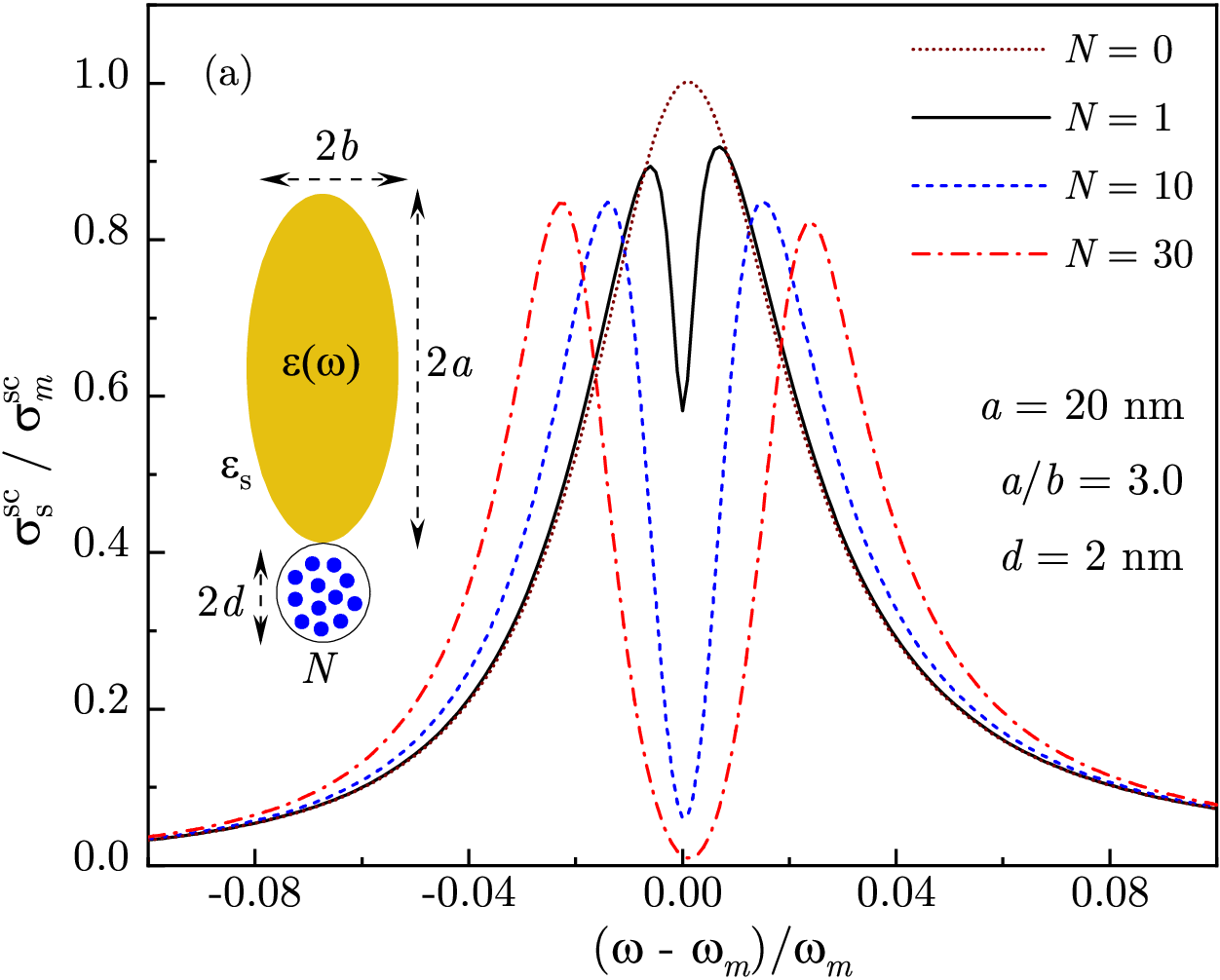}

\vspace{2mm}

\includegraphics[width=0.95\columnwidth]{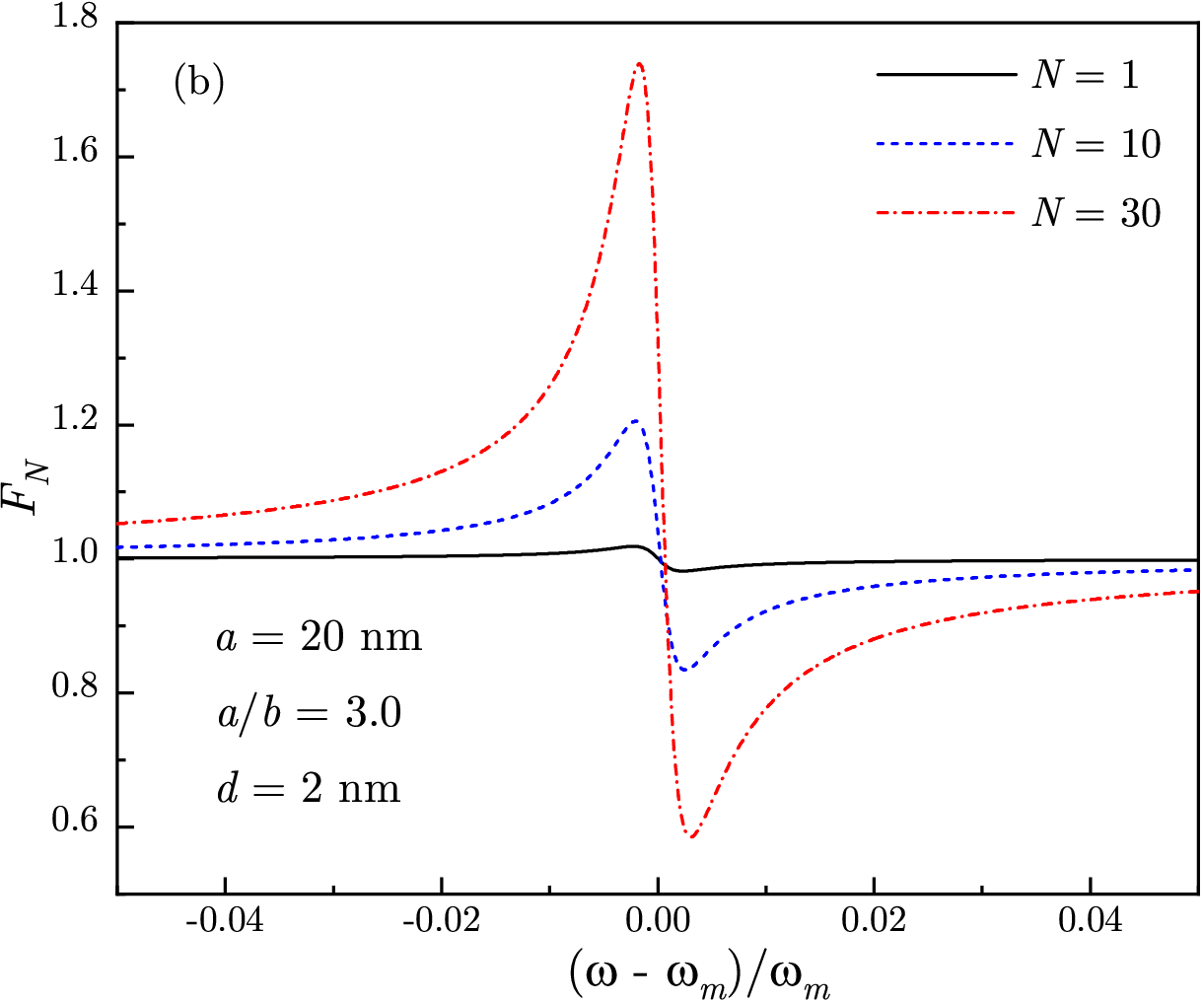}
\caption{\label{fig4} Normalized system scattering cross-section (a) and Fano function (b) are shown for nanorod length 40 nm at several values of QE number $N$. Inset: schematics of a hybrid plasmonic system with $N$ QEs distributed within a spherical region of diameter $2d=4$ nm near the tip of an Au nanorod  with aspect ratio $a/b=3.0$.
 }
\end{center}
\vspace{-3mm}
\end{figure}
%

%
\begin{figure}[tb]
\begin{center}
\includegraphics[width=0.95\columnwidth]{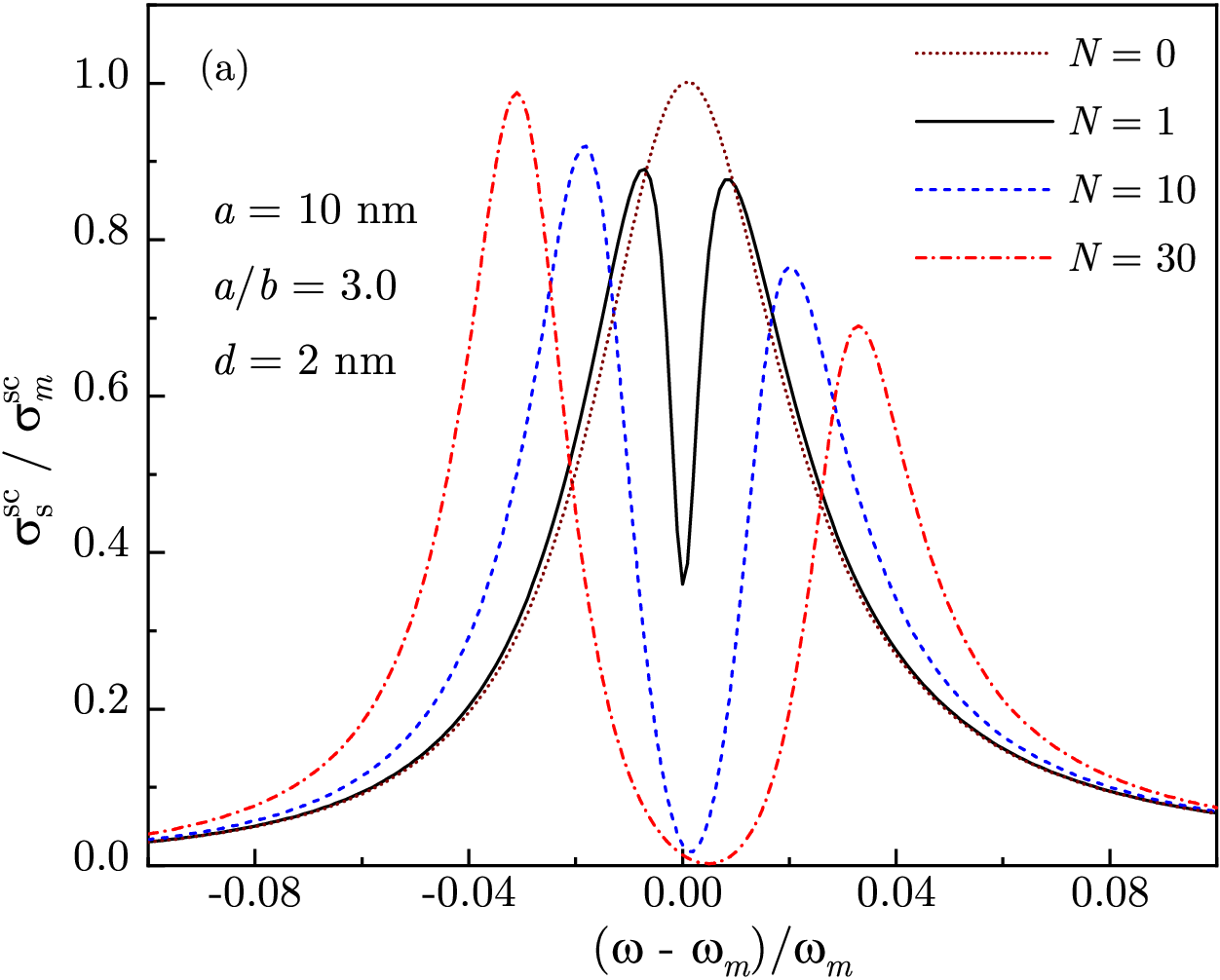}

\vspace{2mm}

\includegraphics[width=0.95\columnwidth]{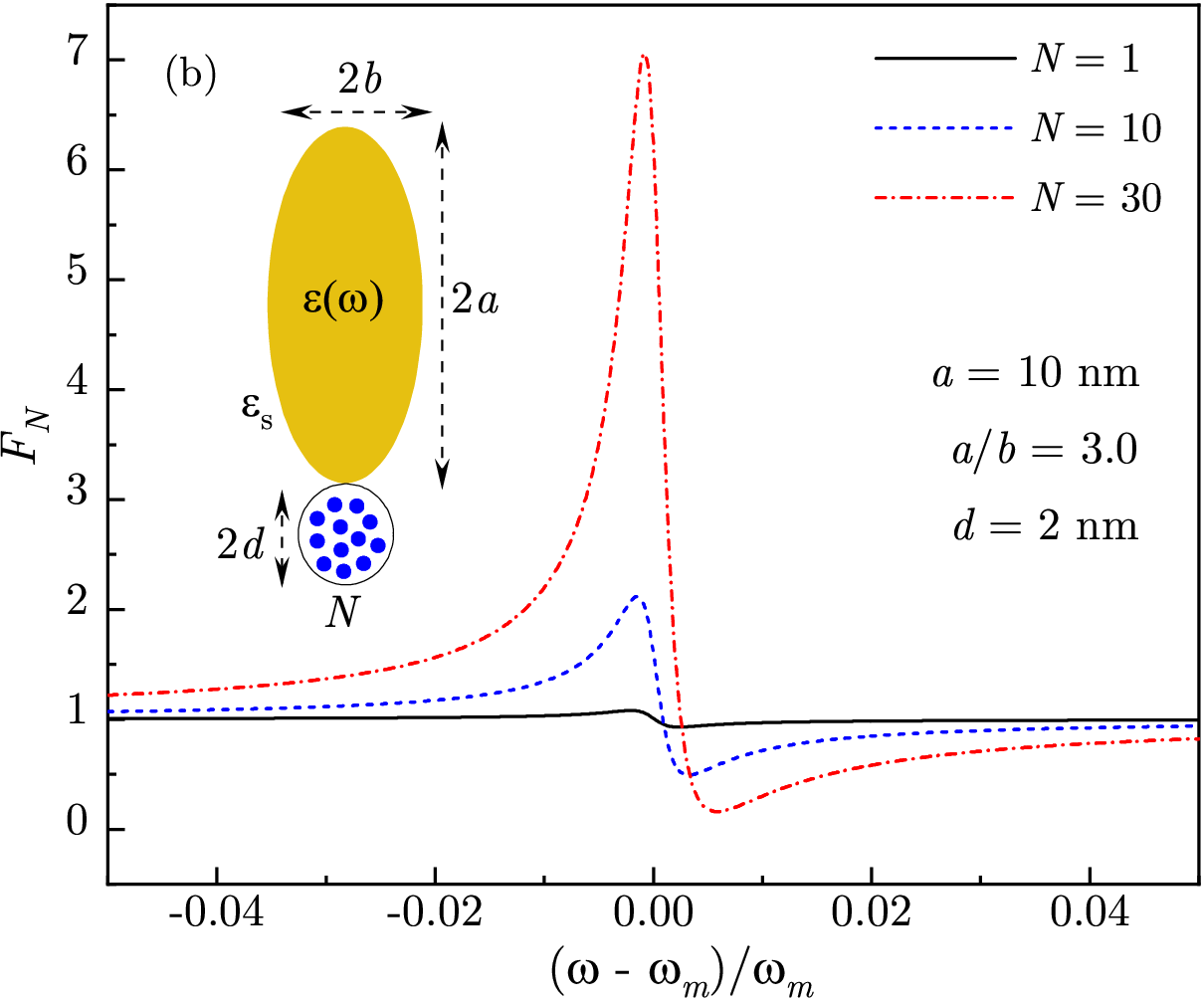}
\caption{\label{fig5} Normalized system scattering cross-section (a) and Fano function (b) are shown for nanorod length 20 nm at several values of QE number $N$. Inset: Schematics of a hybrid plasmonic system with $N$ QEs distributed within a spherical region of diameter $2d=4$ nm near the tip of an Au nanorod  with aspect ratio $a/b=3.0$.
 }
\end{center}
\vspace{-3mm}
\end{figure}
%

A striking feature of the bright state spectra is a sharp increase in their amplitude with $N$, which is caused by the bright state's cooperative coupling with the EM field. Since the corresponding optical matrix element scales as $\mu_{b}^{2}\propto N$, as discussed above, the scattering cross-section accordingly scales as  $\sigma_{bs}^{sc}\propto \mu_{b}^{4}\propto N^{2}$. However, in full spectra, this rapid amplitude rise is masked due to a  much larger amplitude of the dressed plasmon contribution (see below) and, therefore, it can be observed only for very large QE numbers $N$.

A common feature for the scattering spectra shown in Figs.~\ref{fig2} and \ref{fig3} is their distinct spectral asymmetry as the upper-energy polaritonic band is visibly enhanced. Similar asymmetry is also predicted in the coupled oscillators model \cite{pelton-oe10}, widely used for modeling strongly-coupled systems, as it stems from the fact that the higher energy states radiate at a higher rate $\propto \omega^{3}$. In fact, this generic spectral asymmetry is inherent to systems interacting with the EM field as a single dipole, either dressed plasmon's or bright collective state's one. Below we show that the Fano interference between them combined with the plasmon-induced coherence can dramatically change the shape of scattering spectra.

In Figs.~\ref{fig4} and \ref{fig5}, we plot the full scattering spectra of the hybrid plasmonic system, calculated with $\bm{p}_{s}$ given by Eq.~(\ref{dipole-system2}), along with the corresponding Fano functions for larger and smaller nanorods, respectively. Note that, for relatively low QE numbers used here, the contributions of $\bm{p}_{bs}$ and $\bm{p}_{d}$ to the full spectra are negligibly small (not visible in the presented figures). Therefore, it is suitable to interpret the calculated full spectra in terms of dressed plasmon scattering spectra modulated by the Fano function [see Eq.~(\ref{scattering-fano})].

In Fig.~\ref{fig4}(a), we show the evolution of normalized scattering spectra for an $a=20$ nm nanorod with increasing QE number $N$ along with the plasmon scattering spectrum ($N=0$). For a single QE, the system is in the weak coupling regime as the spectrum shows a narrow ExIT minimum within the broad plasmon band. Note that, for $N=1$, the upper polaritonic band is relatively enhanced, similar to  dressed plasmon spectra [see Fig.~\ref{fig2}(a)], althought the spectral asymmetry is somewhat weaker. With increasing $N$, as the system transitions to the strong coupling regime, the two polaritonic bands become nearly symmetric ($N=10$), and with further increase in $N$, the main spectral weight  shifts to the lower polaritonic band ($N=30$), in sharp contrast with dressed plasmon spectra [see Fig.~\ref{fig2}(a)]. Such an \textit{inversion} of spectral asymmetry can be traced to the Fano function, shown in Fig.~\ref{fig4}(a), which modulates the scattering spectra. For $N=1$, the Fano parameter $q_{N}$ is relatively small, so that the Fano function (\ref{fano-function}), while being asymmetric, shows only a weak variation near the value $F_{N}=1.0$ and, hence, does not significantly affect the scattering spectrum. With the increasing QE number, the Fano parameter scales linearly with $N$ [see Fig.~\ref{fig1}(a)], resulting in a larger amplitude of the Fano function [see Fig.~\ref{fig4}(b)], which now  suppresses the higher frequency region while enhancing the lower frequency region. As a result, the asymmetry of dressed plasmon scattering spectra, evident in Fig.~\ref{fig2}(a),  is now inverted [see Fig.~\ref{fig4}(a)]. We stress that the linear scaling of the Fano parameter $q_{N}$ with the QE ensemble size $N$ is a \textit{cooperative effect} caused by the plasmon-induced coherence of individual QEs comprising the bright collective state.

In Fig.~\ref{fig5}, we show the scattering spectra and the Fano function for a smaller nanorod ($a=10$ nm) but otherwise same parameter values. It is evident that a stronger field confinement leads to a greater role of Fano interference as the plasmonic antenna size is reduced. In this case, even for a single QE ($N=1$), the main spectral weight rests with the lower polaritonic band, and it further increases for larger QE numbers [see Fig.~\ref{fig5}(a)]. This behavior can be traced to the evolution of the Fano function with $N$  which shows a dramatic increase in its amplitude [see Fig.~\ref{fig5}(b)]. To understand this behavior, we note that, for large  $q_{N}$, the maximal and minimal values of $F_{N}$ can be found to scale as $q_{N}^{2}$ and $1/q_{N}^{2}$, respectively, implying that, for large $N$, the amplitude of the Fano function scales as $N^{2}$ as well, which is yet another manifestation of the cooperative effect discussed here.

Finally, for relatively low QE numbers studied here, our numerical calculations did not reveal any significant effect of dark states, described by  Eq.~(\ref{dipole-dark}), on calculated scattering spectra. Note that although these states do not couple to the plasmon near field, they can still interact with the EM field that is nearly uniform on the system scale. If, however,  spatial variations of the near field are relatively weak, e.g., in a plasmonic cavity, the dark collective states, which are formed in response to such fields, will couple weakly to the EM field as well. In any case, the dark states play no role in the emergence of polaritonic bands as the system transitions to the strong coupling regime and manifest themselves as a narrow peak at the QE frequency that is not visible, for the system parameters used, in the scattering spectra presented here.


\section{Conclusions}

In this paper, we developed an analytical model supported by numerical calculations for linear optical response of a hybrid plasmonic system comprised of an ensemble of $N$ emitters resonantly coupled to a  localized surface plasmon in a metal-dielectric structure with a characteristic size below the diffraction limit. We have shown that, as the system undergoes transition to the strong coupling regime,  the  scattering spectra are determined by the interplay between coherent energy exchange, which gives rise to exit ExIT minimum and subsequently to the emergence of polaritonic bands, and  the Fano interference, which affects the spectral weight of the polaritonic bands. The latter mechanism is greatly enhanced for large ensembles due to plasmon-induced coherence as the individual emitters comprising the bright collective state are  driven by same alternating plasmon field and hence oscillate in phase. This cooperative effect leads to linear scaling with $N$ of the Fano asymmetry parameter and to quadratic scaling of the Fano function amplitude, which results in the inversion of spectral asymmetry, predicted by models based upon the dominant role of plasmonic antenna. We also elucidated the mechanism of ExIT for the ensemble of $N$ emitters by comparing the scattering spectra for the dressed plasmon and bright collective state, calculated separately, and found that they are consistent with the energy exchange mechanism of ExIT \cite{shahbazyan-prb20}.

Finally, we considered resonant plasmon coupling with the ensemble of $N$ emitters and made a number of simplifying assumptions that allowed us to obtain our results within the analytical model. First, we have neglected direct dipole-dipole coupling between the emitters by assuming that, typically, the emitters' dipole moments are oriented randomly relative to each other, and so direct dipole coupling between them vanishes \textit{on average} and plays no significant role in the transition. Second, for large ensembles, the system can be described semiclassically with a good accuracy, and so we disregarded the nonlinear effects. Note that our main goal in this paper has been to establish a novel phenomenon of plasmon-induced coherence, which is a cooperative effect originating due to large emitter numbers rather than the nonlinear effects.

\acknowledgments

This work was supported in part by the National Science Foundation Grant Nos.  DMR-2000170, DMR-1856515 and    DMR-1826886.


\clearpage

\end{document}